Here the paper follows

\documentstyle[preprint,aps]{revtex}
\begin{document}
\preprint{IP/BBSR/93-66}
\draft
\title
{\bf \Large {Shape  transition in the epitaxial growth of
gold silicide in Au thin films on Si(111)}}
\author{K. Sekar, G. Kuri, P. V. Satyam, B. Sundaravel,
D. P. Mahapatra and B. N. Dev }
\address
{Institute of Physics, Bhubaneswar-751005, India}
\maketitle
\begin{abstract}

     Growth of epitaxial gold silicide islands
on bromine-passivated Si(111) substrates has been studied by optical
and electron microscopy, electron probe micro analysis and helium ion
backscattering. The islands
grow in the shape of equilateral triangles up to a critical size
beyond which the
symmetry of the structure is broken, resulting in a shape transition
from triangle to trapezoid. The island edges are aligned along
$Si[110]$ directions.
We have observed elongated islands with
aspect ratios as large as 8:1. These islands, instead of
growing along three equivalent [110] directions on the Si(111) substrate,
grow only along one preferential direction. This has been
attributed to the vicinality of the substrate surface.\\
\end{abstract}
\narrowtext
\newpage
       Heteroepitaxy produces strained epitaxial layers due to
lattice mismatch between the substrate and the overlayer. Strained
epitaxial layers have interesting properties and are important in
semiconductor devices \cite{TPP}. Since these layers are
inherently
unstable it is important to understand the mechanism of relaxation of
strain. It has been known for a long time that the formation of
dislocations is a strain relief mechanism \cite{JHvan,JWM}.
However, in recent
years it has been recognized that strained layers are unstable
against shape changes \cite{BJS}. Thus shape changes, such as
island
formation, have been identified as a major mechanism for strain
relief. Recently Tersoff and Tromp \cite{JT} have shown
that strained epitaxial
islands, as they grow in size, may undergo a shape transition. Below
a critical size, islands have a compact symmetric shape. But at
larger sizes, they adopt a long thin shape, which allows better
elastic relaxation of the island stress. They observed such
elongated island growth of $Ag$ on $Si(001)$ surface. Prior to this,
various other groups also observed elongated island growth of $Ge$
on Si(001) \cite{YWM}, $Au$ on $Ag(110)$ \cite{SR}, $Au$ on
$Mo(111)$ and on $Si(111)$ \cite{MM} and $GaAs$ islands on $Si$
\cite{FAP}. In all the aforementioned cases where the elongated
island growth was observed, the deposition was performed under
ultrahigh vacuum (UHV) condition in the submonolayer to a few
monolayers range and on atomically clean substrates.

       Here we present our observation of the growth of
large epitaxial gold silicide islands on Si(111) substrates in a
non-UHV method. Moreover, the epitaxial
structures were obtained from a relatively thick ($\sim$ 1000 $\AA$)
film deposited uniformly on the substrate. After annealing, we have
observed growth of small
equilateral triangular silicide islands reflecting the threefold
symmetry of
the underlying Si(111) substrate. The size of these triangular
islands grow up to a critical size beyond which there is a transition
to a trapezoidal shape. The trapezoidal islands have varying
lengths, but similar widths. The largest of the islands have an
aspect ratio of about 8:1. The main observed features are in
reasonable agreement with the prediction of the recent theory of
shape transition by Tersoff and Tromp \cite{JT}. However, in
this theory dislocation-free strained islands have been considered.
We find evidence for dislocations in the silicide islands as well.
Thus, to our knowledge, we report for the first time the evidence for
both the strain relief mechanisms, namely, the shape transition and
the formation of dislocations in the same system.
Another interesting feature of our
observations is that all the elongated islands are aligned in the
same direction, although one would expect them to grow in three
different orientations imposed by the threefold symmetry of the
Si(111) substrate surface. This additional symmetry-breaking in the
growth of
epitaxial islands has been attributed to the vicinality of the
substrate surface.

The method of sample preparation involves gold
evaporation onto a bromine-passivated \hfil\break Si(111) wafer
(n-type, Sb-doped, 0.005-0.02 $\Omega cm$ ) substrate and subsequent
annealing. The method used for the bromine passivation of the Si(111)
surface is known to provide Br adsorption ($\approx 1/4$ monolayer)
at the atop site on the surface Si dangling bonds on hydrofluoric
acid-etched \cite{JAG} or
cleaved \cite{BND} Si(111) surfaces. Br adsorption inhibits the
surface oxidation process. A detailed x-ray
photoelectron spectroscopic characterization of the Br-treated Si
surfaces has been published elsewhere \cite{KS94}. It was also
shown that a Cu thin film deposited on a Br-passivated Si(111)
surface has interdiffusion behavior very similar to the case where
Cu was deposited on an atomically clean (7$\times$7) reconstructed
Si(111) surface \cite{KS92,KS93}.  It should be noted that chemically prepared
$Br$-passivated $Si(111)$ surface in open air \cite{JAG} and $Br$-adsorbed
atomically clean $Si(111)$ surface under ultrahigh vacuum condition
\cite{PFunke} show identical behaviour regarding $Br$-adsorption
site, $Si-Br$ bond length, substrate surface relaxation etc. Also the
presence of some impurities on the $Si$ substrate prior to metal
deposition is not an impediment for epitaxial silicide growth. In
this case the concept of self-cleaning interface has been discussed
by Lau and Mayer \cite{Lau}. Epitaxial growth on $Br$-passivated
substrate assumes futher importance in the light of the recent
spurt of activities on impurity (surfactant)-controlled
epitaxial growth \cite{DJE}. In the present work a  1200 $\AA$ thin Au film
was evaporated from a W basket onto a Br-passivated Si(111)
substrate
at room temperature in high vacuum (10$^{-6}$ Torr). Then the sample
was annealed at (360$\pm 10)^{o}$C, that is
around the $Au-Si$
eutectic temperature of 363$^o$C \cite{AAJ}, for 20 minutes in high vacuum.
This gave rise to the triangular and trapezoidal island structures
shown in Fig. 1 and Fig. 2.  The triangular islands have strong
similarities with the growth of
(7$\times$7) reconstructed triangular domain growth on an atomically
clean Si(111) surface in the low temperature phase of the
($7\times 7)
\leftrightarrow (1\times 1$) order-disorder transition
\cite{WT}.
We also observed fractal structures of $Au$
for this system which will be presented elsewhere \cite{KSfrac}. Longer
(50 min and 80 min) annealing, though changes the $Au$ fractal
structure, does not have any significant effect on the large silicide
island structures.

    In Fig. 1, we notice the formation of the equilateral triangular
structure reflecting the three-fold symmetry of the Si(111) surface.
This indicates that these triangular islands are crystalline $-$
epitaxially grown on the substrate. The composition of the
triangular structures, as determined from electron
probe micro analysis (EPMA), is $Au_{4}Si$ [actually (79$\pm$ 3)$\%
Au$, (21 $\pm$ 3)$\% Si$ for both
triangular and trapezoidal islands]. Here we observe
equilateral triangles of almost identical size. The edges of the
equilateral triangles are aligned along substrate [110] directions.
This alignment has also been observed for silicide growth on a
$Si(111)$ surface under UHV condition \cite{MM} as well as for gold
silicide precipitates in the bulk \cite{FHB}.
In Fig. 2, we observe some equilateral
triangular islands which have grown larger, but mostly we notice
islands of trapezoidal shape with larger areas than that of the
triangles.  The widths of all the trapezoidal islands are comparable
and are roughly equal to the width
of the largest equilateral triangular island. That is, growth of the
islands beyond a critical size, represented by the largest triangular
island, is associated with the shape transition: triangle to
trapezoid.

      We will discuss our results in the light of the recent
theory given by Tersoff and Tromp \cite{JT}, who treated
the case of $Ge$ growth on $Si(100)$ as a generic case.  For a
strained island on a substrate
they computed the energy of an island as a sum of two contributions
$-$ one from the relevant surface and interface energies and the
other from elastic relaxation. The surface energies are
those of the substrate and of the island top and edge facets, and
the interface energy is that for the island-substrate interface.
The
second contribution arises from the fact that an island under stress
exerts a force on the substrate surface, which elastically distorts
the
substrate. This lowers the energy of the island at the cost of some
strain in the substrate. Tersoff and Tromp derived an expression for
the energy per unit volume (E/V) of a rectangular strained epitaxial
island :

\begin{equation}
{\frac {E}{V}} = 2\Gamma (s^{-1}+t^{-1}) - 2ch [s^{-1}
ln[{\frac {s} {\phi h}}] + t^{-1} ln[{\frac {t} {\phi h}}]]
\end{equation}

where $s$, $t$ and $h$ are width, length and height of the island,
respectively; $\phi = e^{-3/2}cot\theta,  \theta $
being the contact
angle; $\Gamma$ contains the surface and interface energies; $c$
involves the bulk stress in the island and the Poisson ratio and
shear modulus of the substrate.
It is clear from Eq.(1) that the surface energy dependent
term
prefers to have a large area island for stability. On the other hand
the strain relaxation energy term prefers to have
islands of smaller area for greater stability. The optimal tradeoff
between surface
energy and strain is obtained through the minimization of E/V with
respect to
$s$ and $t$. This gives $s = t = \alpha_o$, where

\begin{equation}
\alpha_o = e \phi h e^{\Gamma/ch}
\end{equation}

For island sizes $s, t < e\alpha_o$, the square island shape ($s=t$)
is stable. Once the island grows beyond its optimal diameter
$\alpha_o$ by a factor of $e$, the square shape becomes unstable and
a transition to rectangular shape takes place. As the island grows,
the aspect ratio $t/s$ becomes ever larger.

     The abovementioned treatment was for epitaxial islands on a
Si(100) surface which has a four-fold symmetry.
In this case the island growth up to the critical size has a four-fold
symmetry (square). Beyond the critical size the islands grow
in rectangular shape. The long rectangular
islands have been called self-assembling quasi-one dimensional
``quantum wires".
In our case the substrate, $Si(111)$, has three-fold symmetry.
Therefore,
the islands up to the critical size are of equilateral triangular
shape and the shape transition is from
triangle to trapezoid. Here in the initial stage of the growth the
triangular islands are of submicron size and may be called
self-assembling quasi-zero dimenional ``quantum dots".

    The thickness of the $Au_{4}Si$ islands is $\>^\sim 1\mu m$,
which is estimated from the penetration depth of 10 and 25 keV
electrons used for the EPMA composition analysis.
The tree-like structures in Fig. 1 consist of Au.

      At this point let us try to make some quantitative estimates
from the triangular and the trapezoidal islands observed in our
experiments. Fig. 3(b) shows the plots of $l_1$ and $l_2$ versus
island
area $A$. It is seen that the islands grow as equilateral triangles
($l_1=l_2=l_o$) up to a critical size beyond which there is a transition
to trapezoid ($l_2>l_1$). For the largest triangular islands the
area is ${\sqrt 3} l_o^2/4 = 110 \mu m^2$ (measured). In the light
of Tersoff and Tromp theory, we attempt to give an
approximate estimation of the energy per unit volume ($E/V$). We set
the area of the island, where the shape transition takes place,
$e^2 \alpha_o^2 = 110 \mu m^2$. This provides $\alpha_o = 3.86 \mu m$.
With $h=1 \mu m$ and assuming $\theta = 45^o$, we get
$\Gamma/ch = 1.85$. For the trapezoidal islands we use
$s={\sqrt 3}  l_1/2$ and $t=l_2 - s/{\sqrt 3}$ ($A = s t$)
and use Eq. (1)
to evaluate $E/V$. For the triangular islands we use $s={\sqrt 3}
l_1/2$ and $t=l_2/2$ ($A = s t$) to evaluate $E/V$. The results
are shown in Fig. 3(a) \cite{star}. Our results are in reasonable
agreement with the gross features of the Tersoff and Tromp theory.
However, we do not observe the sharp change in the aspect ratio
around the transition point as predicted by the theory. If the
second derivative of the energy with respect to the island size is
discontinuous as mentioned in Ref.\cite{JT}, the magnitude of the
discontinuity must be too
small to be detected in the present experiment.

   In the low energy electron microscopy (LEEM) study of gold silicide
growth on $Si\{111\}$ for submonolayer $Au$ deposition, Mundschau
et al. \cite{MM} observed triangular island growth for low
coverages and
rod shaped island growth at higher coverages with the major axis of
these rods aligned along $Si[110]$ directions.  The rod-like islands
had
an aspect ratio of $\sim 8:1$. However, in their study
they found the elongated islands to be
aligned along the three [110] directions on $Si\{111\}$ as expected
from the symmetry. Our observation of unidirectional elongated island
growth along the $[0{\bar 1}1]$ direction may be explained from the fact
that our substrate surface was vicinal, i.e., the $(111)$ surface was
misoriented by $4^{o}$ towards $[2\bar 1 \bar1]$ azimuthal direction.
This misorientation gives rise to the formation of single-layer and
triple-layer steps which run along the $[0\bar 1 1]$ direction \cite{EPL}.
Thus our observed elongated islands are along the length of the
steps. It is well known that nucleation and growth are predominant
at surface defects. In our case the growth might have
occured preferentially at the steps. Vicinal surfaces of
semiconductors are usually used as substrates for epitaxial
overgrowth (such
as directed epitaxy) and new interesting electronic properties of
these systems have been predicted \cite{PMP}.

   Rutherford backscattering of 2 MeV $He^{+}$ ions from the
sample provided evidence for the presence of a thin ($\sim$ 50 $\AA$)
layer containing gold even in the island-free region (the apparently
depleted flat region in Fig. 2). From EPMA
analysis
of this region we could not determine whether this $Au$ exists as
$Au_{4}Si$, because
the high energy electrons penetrate much deeper into
the sample and leads to overestimation of the $Si$ content. For $Au$
deposition (100-1000 $\AA$) on $Si$ at room temperature, followed by
annealing at $T \leq 400^{o}C$, and studied by other techniques, such
as AES, a continuous silicide layer thickness of $\sim 30  \AA$ was
previously reported \cite{GLL}. In our optical color micrograph
the flat region and the islands appear to be of the same
color which points to a composition of $Au_4Si$ for the flat region
as well. If this thin $Au_{4}Si$ layer is epitaxial, it would amount
to a layer-plus-island or Stranski-Krastanov growth of $Au_{4}Si$.
The composition of the stringy structure in Fig. 2 as determined by
EPMA is $100 \%$ $Au$.

      The detailed features of the
$Au_{4}Si$ islands are not visible in the optical micrographs shown
in Fig. 1 and Fig. 2. A scanning electron micrograph is shown in
Fig. 4. The patterns with pin-hole structures on the islands
indicate the
presence of dislocations. In fact, in some cases we observed porous
structure of the islands. Thus, in addition to the mechanism of shape
transition for strain relief there is a partial strain relief through
the formation of dislocations in these strained epitaxial $Au_{4}Si$
islands. As
Tersoff and Tromp \cite{JT} point out, a partial strain relief
through dislocations would not affect the general aspects of shape
transition except for reducing the effective value of bulk stress of
the island material. One would, of course, expect to see a larger
critical size for islands with higher dislocation density. This might
be partly responsible for the large size ( area $\sim$1000 times
those observed in Ref.8) of the gold silicide islands on the
bromine-passivated Si(111) surface.

       On the Si(111) substrates with a thin oxide layer we did not
observe epitaxial island growth upon $Au$ deposition and subsequent
annealing at 360$^o$ C. It has been previously shown that
the diffusion behavior is
different for a metal layer deposited on a bromine-treated Si(111)
substrate and on an Si(111) substrate with a native oxide layer, the
bromine-treated substrate behaving like an atomically clean
substrate and the native oxide acting as a diffusion barrier at the
interface \cite{KS92}.

       In conclusion, growth of epitaxial structures of
$Au_{4}Si$ has been observed
in Au thin films prepared by vacuum evaporation of Au
on bromine passivated Si(111) substrates and subsequent vacuum
annealing at 360$^{o}$C. These structures reflect the threefold
symmetry
of the underlying Si(111) substrate. The epitaxial structures are of
equilateral triangle in shape. These epitaxial triangular islands
grow
bigger up to a certain critical size.  Beyond the critical size the
triangular structures undergo a shape transition to an elongated
trapezoidal shape. All the elongated islands are aligned in the
same direction, although one would expect them to grow in three
different orientations imposed by the threefold symmetry of the
Si(111) substrate surface. The threefold symmetry in the elongated
island
growth is apparently broken due to the vicinality of the substrate
surface. We observed islands with aspect ratios as large as 8:1.
However, under appropriate conditions,
the islands may grow much longer in the preferential direction. In
the scanning electron micrograph of the islands, we
found evidence for dislocations. This implies, both the strain relief
mechanisms $-$ the shape transition and the formation of
dislocations $-$ are
concomitant in the growth of $Au_{4}Si$ islands on a
bromine-passivated Si(111) substrate. This may be, after all a
general feature in the growth of strained epitaxial islands.

\acknowledgements

  We thank Prof. Pham V. Huong and Dr. M. Lahaye
for some of the
EPMA measurements and Dr. B. K. Mohapatra for taking the scanning
electron micrograph.

\newpage
\begin{figure}
\caption{ Photomicrograph showing  $Au_{4}Si$ islands, that is,
epitaxial $Au_{4}Si$ crystallites on the
Si(111) substrate. The three-fold symmetry is reflected in
the shape of many islands. Smaller structures are formed for shorter
duration of annealing. Diameter of the image is $400 \mu m$.}
\end{figure}
\begin{figure}
\caption{ $Au_{4}Si$ islands in the shape of trapezoid.
Although the islands vary in length, their width remains practically
constant and approximately equal to that of the largest triangular
islands. Diameter of the image is $400 \mu m$. }
\end{figure}
\begin{figure}
\caption{  (a) Computed energy per unit volume of island, in
units of $ch/ \alpha_o$, vs measured island area $A$ ($\circ $- triangle,
$\star$- trapezoid). The solid line is a polynomial fit.
Beyond the transition point if the islands
remained triangular the energy would be slightly (only 7$\%$ for the
largest island) higher. \hfil\break
(b) Lengths $l_1$ and $l_2$ (as shown in the
inset) of the islands, vs $A$.
Unit of length for $l_1$ and $l_2$ is
$\alpha_o$ and that of $A$ is $\alpha_o^2$ [see text and Equ.(2)].
Solid lines are to guide the eye. For $A < $ $e^2
\alpha_o^2$, $l_1$ and $l_2$ are equal (equilateral triangle).}
\end{figure}
\begin{figure}
\caption{ A scanning electron micrograph showing pin-hole
patterns in the strained $Au_{4}Si$ islands. The length of the middle
island is $\approx 20 \mu m$.}
\end{figure}

\begin{references}
\bibitem{star} For the triangular islands, assumption of $s = t =
{\sqrt A}$ yields $E/V$ values which deviate by $\sim 1\%$ from the
curve shown in Fig. 3(a).
\end{references}
\end{document}